\documentclass[12pt]{iopart}

\usepackage{graphicx}
\usepackage{iopams}  
\usepackage{url}
\usepackage[dvipsnames]{color}

\begin{document}

\title[nEoS: a neutron-star EoS interpolation]{nEoS: Neutron Star Equation of State\\ from hadron physics alone}

\author{Eva Lope Oter$^{1*}$, Andreas Windisch$^{2,3}$, Felipe J. Llanes-Estrada$^1$ and Mark Alford$^2$}

\address{$^1$ Depto. de F\'{\i}sica Te\'orica, Univ. Complutense de Madrid, Plaza de las Ciencias 1, 28040 Spain}
\address{$^2$ Department of Physics, Washington University in St. Louis, MO, 63130, USA}
\address{$^3$ Department of Physics, University of Graz, Universit\"atsplatz 5, 8010 Graz, Austria}
\address{$^*$ Contact author.}
\ead{mariaevl@ucm.es}

\begin{abstract}
We contribute a publicly available set of tables and code to provide Equations of State (EoS) for matter at neutron star densities. Our EoSes are constrained only by input from hadron physics and fundamental principles, without feedback from neutron star observations, and so without relying on General Relativity. They can therefore be used to test General Relativity itself, as well as modified gravity theories, with neutron star observables, without logical circularity. We have adapted state of the art results from $NN$ chiral potentials for the low--density limit, pQCD results for the asymptotically high--density EoS, and use monotonicity and causality as the only restrictions for intermediate densities, for the EoS sets to remain as model--independent as is feasible today. 
\end{abstract}

\vspace{2pc}

%
\submitto{\JPG  \\ Focus Issue on Hadrons and Gravitational Waves after GW170817}
%
%
%

\newpage
\section{Introduction}

Neutron stars have long offered possible tests of General Relativity at the energy density of nuclear matter~\cite{Dobado:2011gd,Doneva:2015hsa,Capozziello:2015yza}, but the discovery of gravitational waves has accelerated the pace of activity~\cite{Loutrel:2017fgu,Resco:2016upv}, and the multimessenger observation of neutron star inspirals~\cite{TheLIGOScientific:2017qsa} has opened an entire new field~\cite{Chatziioannou:2015uea}.

In relating neutron star measurements such as masses, radii, vibrational modes or gravitational wave emission to General Relativity (GR), a key component is the Equation of State (EoS) of the material in the neutron star.

A broadly held view is that hadron theory is hopelessly difficult and that the Equation of State of neutron stars will eventually be determined or at least strongly constrained from neutron star observations~\cite{Chatziioannou:2015uea}. This would be akin to other instances, such as the determination of parton distribution functions, where theory computations, with strong model assumptions that thwart control of systematic uncertainties, are substituted by direct use of high--energy data--based  extractions in predicting cross sections.

However, our point here~\cite{Llanes-Estrada:2016oom} is that an EoS determined this way cannot be used to constrain theories beyond General Relativity because it already assumes GR (in its use of Tolman-Oppenheimer-Volkoff equations to calculate an observational mass--radius diagram, or in its use of the GR gravitational wave solutions, for example). To avoid circular reasoning, the community needs an effort to provide EoS purely from hadron theory. The purpose of this brief article is to report an endeavor in this direction that we have carried out and now put at the disposal of other researchers. 

In section~\ref{sec:input} we collect and analyze references to what, in our opinion, currently forms the most reliable source of knowledge on the EoS from first-principles hadron theory. This is in the first place perturbative Quantum Chromodynamics (pQCD) for asymptotically high density (presumably valid outside the range of densities where neutron stars are sensitive to the EoS, but constraining the EoS nonetheless) and Chiral Perturbation Theory (ChPT) for low and moderate densities where neutrons in medio behave not too differently from neutrons in vacuo.

In section~\ref{sec:interpolate} we briefly show how the basic principles of causality and monotonicity of the EoS help constrain it through the intermediate density regime most relevant for neutron star properties. Section~\ref{sec:public} then describes our wrapper 
to put this information within reach of Numerical Relativity practitioners and others who
may want to test General Relativity or modifications thereof with neutron stars (some tests have been run and reported in Section~\ref{sec:tests}). Our conclusions 
and suggestions for further progress are then spelled out in section~\ref{sec:conclusions}.

To summarize the outcome of the work: we provide a set of Equations of State that sample the uncertainty bands of Chiral Perturbation Theory (at low density), likewise sample the band of perturbative QCD (at high density) and interpolate between both at intermediate density with the only restrictions of causality and monotonicity.

\newpage
\section{State of the art computations} \label{sec:input}

Historically, the $\beta$--equilibrated (mostly neutron) matter EoS for neutron stars  has been obtained by 
feeding ``realistic'' nuclear potentials into the many body methods appropriate
for continuous matter~\cite{Akmal:1998cf}. In the last decades, however, a new family of potentials
has appeared, based on modern chiral interactions, that promise to improve the precision of 
low--energy QCD. (The LENPIC collaboration~\cite{Binder:2018pgl} has demonstrated that the spectroscopy of 
light nuclei can be successfully treated, so that a first--principles foundation for nuclear physics
is ever closer). 
Such improvement is naturally applicable to the low--density EoS, and because the interactions are derived from QCD through effective theory, this makes it possible 
for the first time to obtain systematic uncertainties in the obtained EoS.

Among the chiral potentials, we have consulted several contemporary works \cite{Sammarruca:2014zia,Hu:2016nkw,Holt:2016pjb,Drischler:2016djf,Hebeler:2015wxa}
and extracted the EoS (pressure $P$ as function of the energy density $\varepsilon$) from their data via
\begin{eqnarray}
\varepsilon= n\left(M_N+\frac{E}{A}\right)\\ 
P= n^2\,\frac{d(E/A) }{d n}\ .
\end{eqnarray}
Here, $E/A$ is the energy per nucleon in nuclear matter, $n$ is the nucleon--number density, and $M_N$ the nucleon mass. Chiral perturbation theory is used only up to an energy density of order 200 MeV/fm$^3$  ($0.2$ GeV$^4$), so that the nucleons are not very relativistic (an exception is the work of~\cite{Hu:2016nkw} that extends to 375 Mev/fm$^3$).

It is worth noting that chiral perturbation theory follows the systematic counting of nuclear--force terms, described for example in~\cite{Meissner:2007zza}. 
In these implementations, the precise counting in vacuo has been employed: assessments based on the counting modified for nuclear matter~\cite{Lacour:2009ej} are still needed~\cite{Oller:focus} and it would be interesting to  have systematic studies thereof. With this caveat in mind, we believe that the current status of the calculations already allows us
to set reasonably systematic uncertainty bands to low--energy predictions.

The EoS in our collection use Chiral Perturbation Theory with baryons. We use ChPT calculations that go at least to NLO and reach in some cases up to N$^4$LO.  The higher order sets typically treat the two--body nucleon--nucleon interaction properly but use an approximation to the three-body force (or include it only at the lowest order). We acknowledge that this causes a systematic bias, lowering the repulsion of the nuclear interaction and thus the stiffness of the EoS (since it is known that three--body forces lower the attraction~\cite{Toyokawa:2017pdd}). This is an aspect where systematic improvement is possible in the next few years. Further work by specialists should reaffirm our trust in the low-density uncertainty band, and address the systematic error introduced by simplifying three and many-body forces, which is making current EoS a tad too soft.

Further, the chiral interactions are cut off at a momentum transfer where the chiral expansion deteriorates and higher resonances or quark-level effects have to be taken into account. The EoS sets available in the literature use cutoffs of 450\,MeV and either 500 or 600\,MeV depending on the reference. In Ref.~\cite{Hu:2016nkw} the long--range part of the interaction is cut off in coordinate space instead of momentum space, and the sensitivity to the cutoff is exposed by providing two sets with $R=0.8$\,m and 0.9\,fm respectively (the short-range contact terms have a standard momentum cutoff $\Lambda=2/R$).

At high density, on the other hand, eventually one enters the regime of pQCD. This has been studied 
in~\cite{Kurkela:2009gj,Kurkela:2014vha} and a new, partial  N$^3$LO calculation has been presented in~\cite{Gorda:2018gpy}~\footnote{In fact, this group~\cite{Annala:2017llu} has, in a study close to our philosophy, interpolated the EoS between their high density computation and a selected low-density one, but incorporating the constraints from neutron star observables. This makes their interpolated EoS less uncertain for conventional astrophysics applications, but not usable to constrain theories modifying General Relativity.}.
For the equation of state at baryon chemical potential above $\mu_{\rm match} \sim 2.6$ to 2.8\,GeV we have made use of the earlier, completed, computations to provide constraints at high density on our sampling of the Equation of State. 

The authors of~\cite{Kurkela:2014vha} provide a handy parametrization of their field theoretical calculation.
The pressure (in GeV$^4$) is given in pQCD, as a function of the baryon chemical potential $\mu_B$ (in GeV) and a parameter $X$ proportional to the renormalization scale ($X\equiv 3 \bar\Lambda/\mu_B$) by the simple expression
\begin{equation} \label{pQCDPressure}
P_{\rm QCD} = \left[ \frac{3}{4\pi^2} \left(\frac{\mu_B}{3}\right)^4\right] \left( 0.9008 - 
\frac{0.5034 X^{-0.3553}}{\mu_B - 1.452 X^{-0.9101}}
\right)\ .
\end{equation}

The first factor between brackets corresponds to the free gas and the second factor stems from the $O(\alpha_s^2)$ corrections.
As recommended by those authors, we have taken as limits for the $X$ parameter the values 1 (low) and 4 (high) to define a band of allowed EoS in the high density regime. The renormalization scale for the two lines that delimit that band (see figure~\ref{fig:Kurkelagraph}) is then, in the $\bar{\rm MS}$, given by $\bar{\Lambda}=\frac{\mu_B}{3}$ and $\bar{\Lambda}=\frac{4 \mu_B}{3}$ respectively.

\section{Interpolation at intermediate densities \\ and construction of the sample EoS in the entire range}\label{sec:interpolate}

The intermediate density range is the most important to determine key properties of neutron stars
such as the maximum allowed mass (known to be about two solar masses~\cite{Demorest:2010bx},
with the highest mass claim~\cite{Linares:2018ppq} currently being $2.27^{+0.17}_{-0.15}M_\odot$).
Yet it is also the region where a consistent expansion of QCD is not workable: we can only resort to basic theoretical properties, causality ($c_s<c$) and monotony ($c_s\geq 0$).

As we do not wish to use polytropic equations of state (assuming such form introduces a systematic bias, and we want to have different systematic uncertainties from other groups working in the field) we have adopted a simple numerical interpolation scheme on a square (logarithmic) grid in the $(\varepsilon,P)$ energy--density/pressure two--dimensional space.

We fix the low--density and high density limits of the intermediate band as specified shortly. At each limit we establish a window of acceptable pressures. At the low density end this is read off from the data $(\rho,E/A)$ in the references mentioned in section~\ref{sec:input}. At the high-density end it is computed from the parametrization of~\cite{Kurkela:2014vha} given in Eq.~(\ref{pQCDPressure}). 
We generate random functions on the grid that interpolate between the small window of maximum pressures for the low-density EoS and the small window of pressures for the high-density EoS. The conditions of causality and monotony translate into 
\begin{equation}\label{constraint}
0\leq \frac{dP}{d\varepsilon} \leq 1\ ,
\end{equation} 
which we simply impose by Von Neumann's rejection method, discarding all generated EoS that do not satisfy this criterion for every $\varepsilon$. We impose \emph{no condition of smoothness nor monotony} to the squared speed of sound (the derivative of the $P(\varepsilon)$ EoS function). This means that the speed of sound can vary dramatically, changing quickly from large
values near 1 (strong repulsive interactions) to low values near zero (which would signal phase transitions, 
broadly expected in cold nuclear matter~\cite{Alford:2015gna,Mellinger:2017dve,Vidana:2018bdi}). Since we are not constraining the number and nature of phase transitions in the intermediate density region, where several alternatives have been proposed (hadron phases with strange hyperons, 2SC condensates, various chiral phases, etc; see for example~\cite{Alford:2015gna} for hybrid stars) we allow our family of EoSes to have widely disparate sound speeds as long as they are consistent with the minimum requirements in Eq.~(\ref{constraint})~ \footnote{Within General Relativity, we know that the EoS will have to be steep, with $c^2_s$ near 1 and definitely above $1/3$ in at least part of this intermediate region~\cite{Chamel:2012ea,Alford:2013aca,Bedaque:2014sqa,Tews:2018kmu}, in order to support $2M_\odot$ neutron stars. 
}.

\begin{figure}
\centerline{\includegraphics[width=5in]{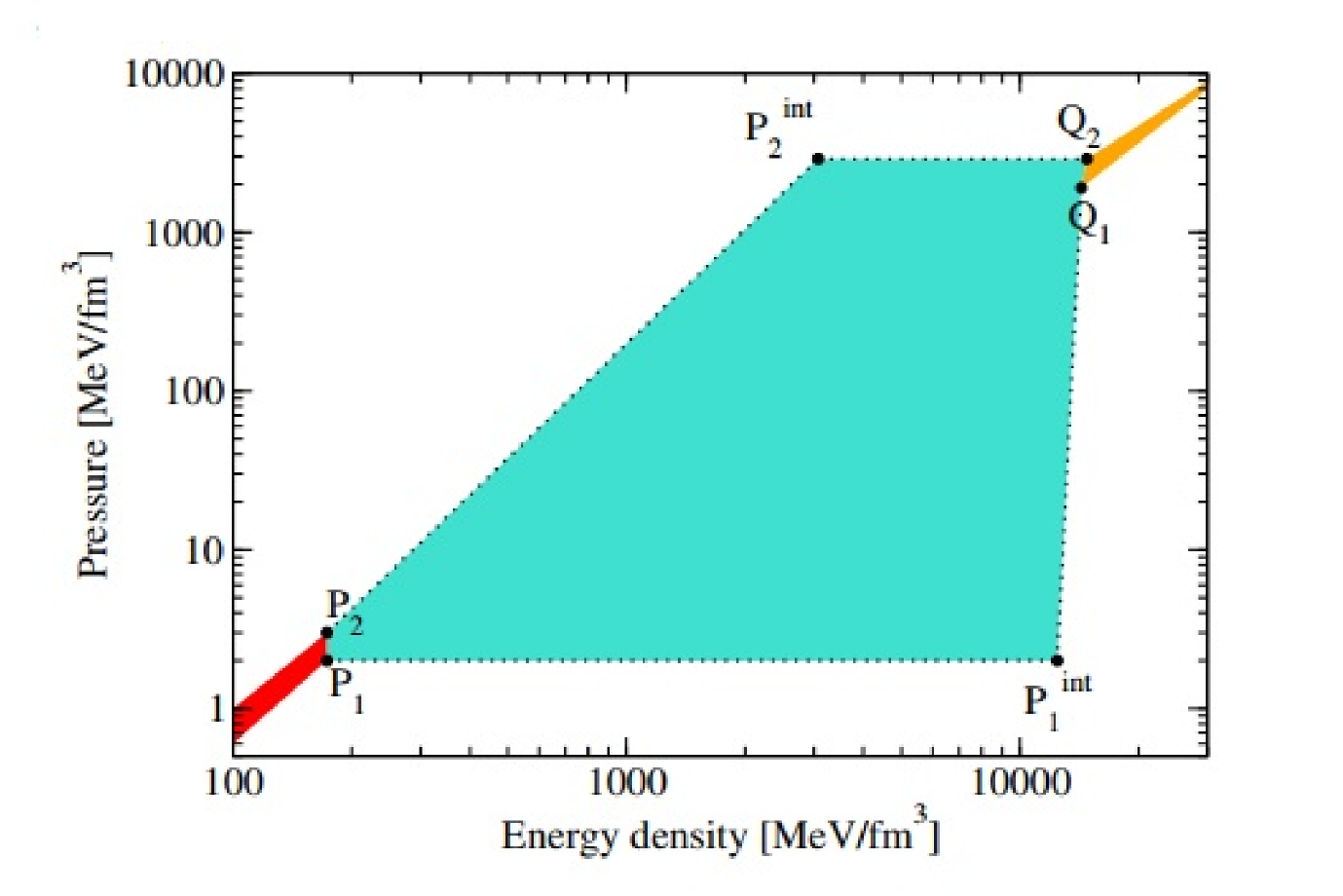}}
\caption{\label{fig:domain} Scheme for the construction of the boundary of the region at intermediate densities (as determined by causality and monotonicity) between chiral perturbation theory (narrow band in the left bottom corner) and high--density physics treated with pQCD (narrow band at the right top corner).}
\end{figure}

\begin{figure}
\begin{center}
\includegraphics[width=0.75\columnwidth]{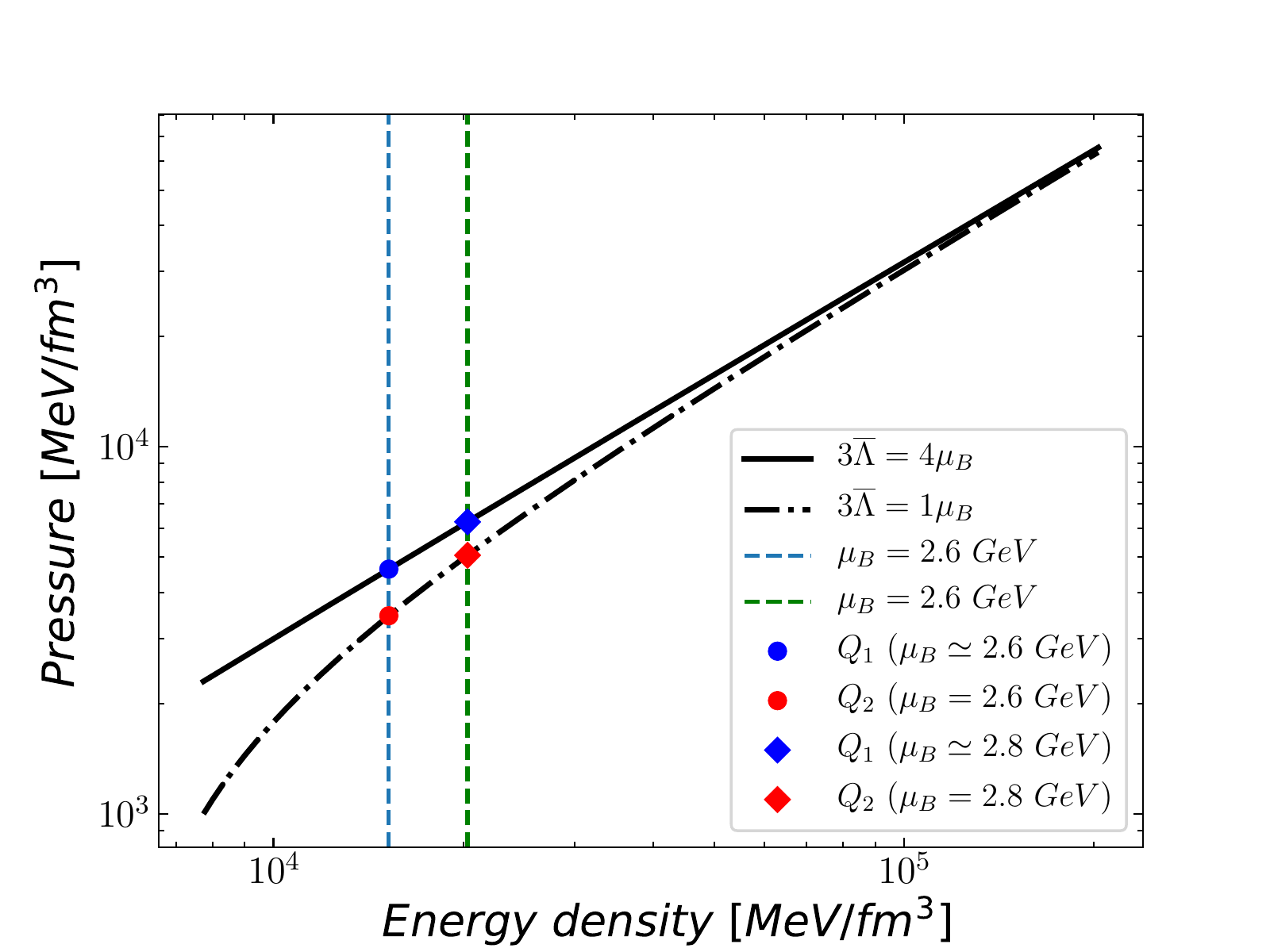}
\end{center}
\caption{
High-density band from pQCD~\cite{Kurkela:2014vha} showing our choice of the points $Q_1$ and $Q_2$ for the two different sample values of $\mu_B$. They are chosen in a relatively safe density range where the pressure  band does not grow very quickly, as happens for lower densities. Since we present $P(\varepsilon)$ and not $P(\mu_B)$, we first fix $Q_2$ from the higher ($X=4$) line with either $\mu_B=2.6$ or 2.8 GeV. We then use the resulting energy density to obtain a slightly modified $\mu_B$ and from Eq.~(\ref{pQCDPressure}) we obtain the low $Q_1$ band end with $X=1$.
Tightening the intermediate density band by lowering this choice of $Q_1$, $Q_2$ would quickly constrain the EoS at intermediate density at a formal level, but much more care would be needed in dealing with the systematic uncertainty introduced. 
\label{fig:Kurkelagraph}
}
\end{figure}

We can easily determine the boundary of the region for which the constraint of Eq.~(\ref{constraint}) can hold true. The construction, that we very briefly describe, is shown in figure~\ref{fig:domain}. Starting at the low--end of the low--density pressure interval,  $P_1$,
we take a line $L^{(0)}_{\rm low}$ of vanishing slope as the first segment of the boundary. The length
of this straight boundary segment parallel to the energy--density axis is determined by its crossing with the steepest possible line $L^{(1)}_{\rm low}$ dropped from the low--end of the high--density pressure window, $Q_1$ in the figure. Algorithmically, we calculate the ($Q_1$,$Q_2$) interval from Eq.~(\ref{pQCDPressure}) as shown in figure~\ref{fig:Kurkelagraph}.
(Physically, the line $L^{(0)}_{\rm low}$  would correspond to a phase transition immediately following chiral perturbation theory, finishing in $L^{(1)}_{\rm low}$ a phase of very repulsive interactions.)  Analogously, the upper density end of the band is determined by a maximum--slope line $L^{(1)}_{\rm high}$ from $P_2$ at low density, followed by a flat line $L^{(0)}_{\rm high}$ to $Q_2$ at high density.

Then, a grid is laid out in that band and for each energy density we choose a random value of the pressure from the values on the grid. The inequalities~(\ref{constraint}) are tested at each point in the grid and if the function fails the test at any point, it is rejected. The remaining EoS form a set that satisfies all constraints. We have executed the program numerous times and stored the resultant set of EoS at the website described in section~\ref{sec:public}.

The low--density band below the $P_1$ and $P_2$ points (these furnishing the starting pressure interval for the interpolation to intermediate densities) in figure~\ref{fig:domain}, is itself also randomly sampled from the lowest density by means of a random walk, imposing only the same very basic conditions. Here pressure grows very slowly between the two lines limiting the band, so exceeding the speed of light is not a concern ($c_s$ is typically 0.01-0.1). But in each step, in addition to not exceeding the side lines of the band, monotonicity of the pressure is also imposed.
Additionally, if a point falls beyond the limiting lines $L^{(i)}_{\rm low}$ or $L^{(i)}_{\rm high}$ of the band it is also rejected. 
Those limiting lines are alternatively taken from each of the low--density sets available in the literature, at fixed order in perturbation theory. However they cover both cutoffs: we simply take a band broad enough that the uncertainty introduced by varying $\Lambda$ (or $R$) between the two quoted values is completely accounted for. If the computed line violates either causality or monotony, it is truncated.

For the high--density band, the procedure is similar. 
We stop the computation of the intermediate energy band upon reaching the line joining the pressures $Q_1$ and $Q_2$ corresponding to an approximate matching baryon chemical potential $\mu_{\rm match}=2.6$ GeV or $\mu_{\rm match}=2.8$ GeV as computed from $Q_2$ (to show sensitivity to the choice) beyond which the energy density is computed by perturbative QCD. 
By construction we end the intermediate band at a pressure that can be matched to the pQCD one. Finally, this last band is also sampled with a random walk from lower to higher densities, as the low--density one.

\begin{figure}

  \begin{minipage}{0.45\textwidth}
   
   \includegraphics*[width=3.44in]{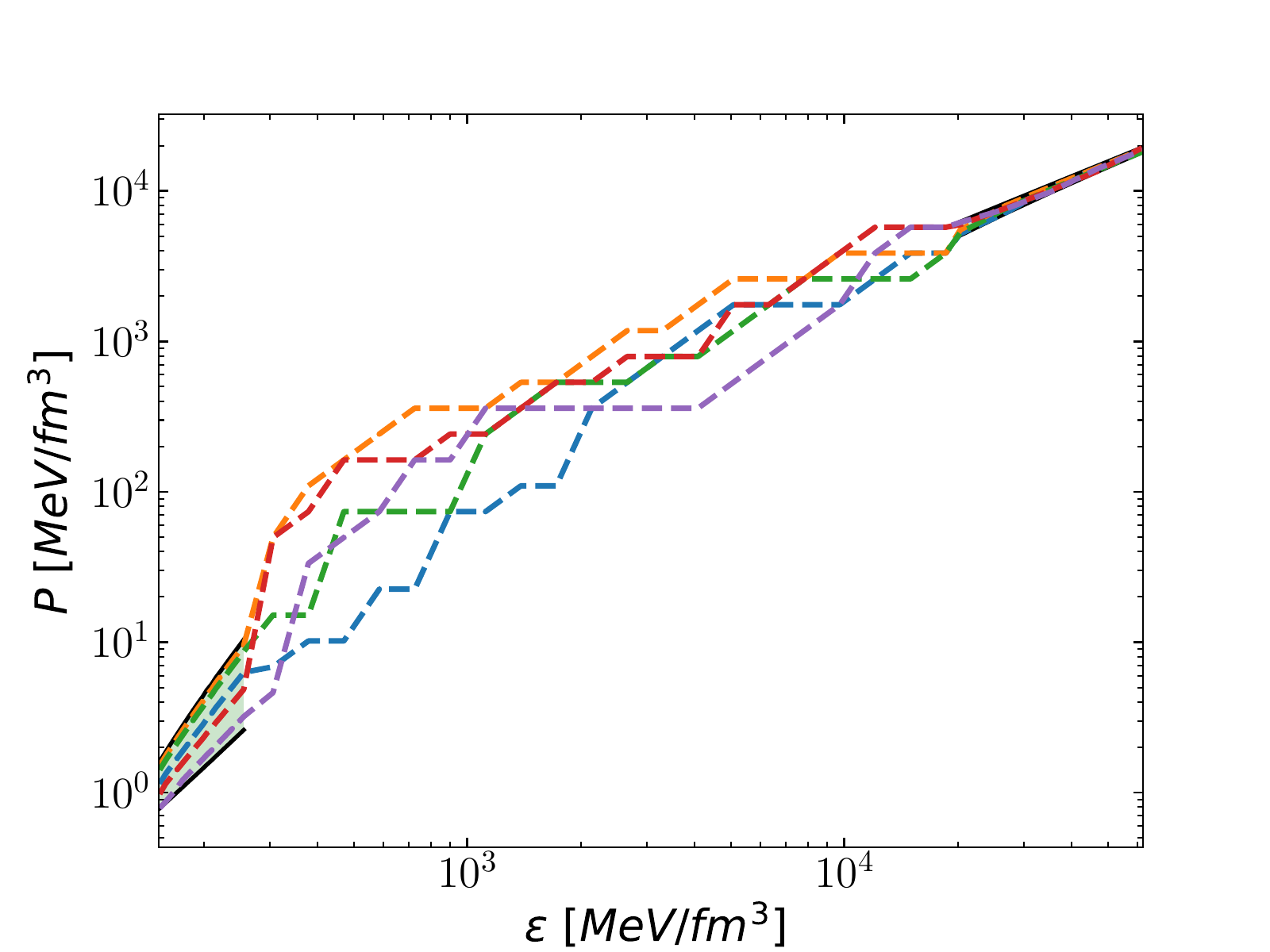}
   \end{minipage}\hspace{0.7cm}
  \begin{minipage}{0.45\textwidth}
   
    \includegraphics*[width=3.44in]{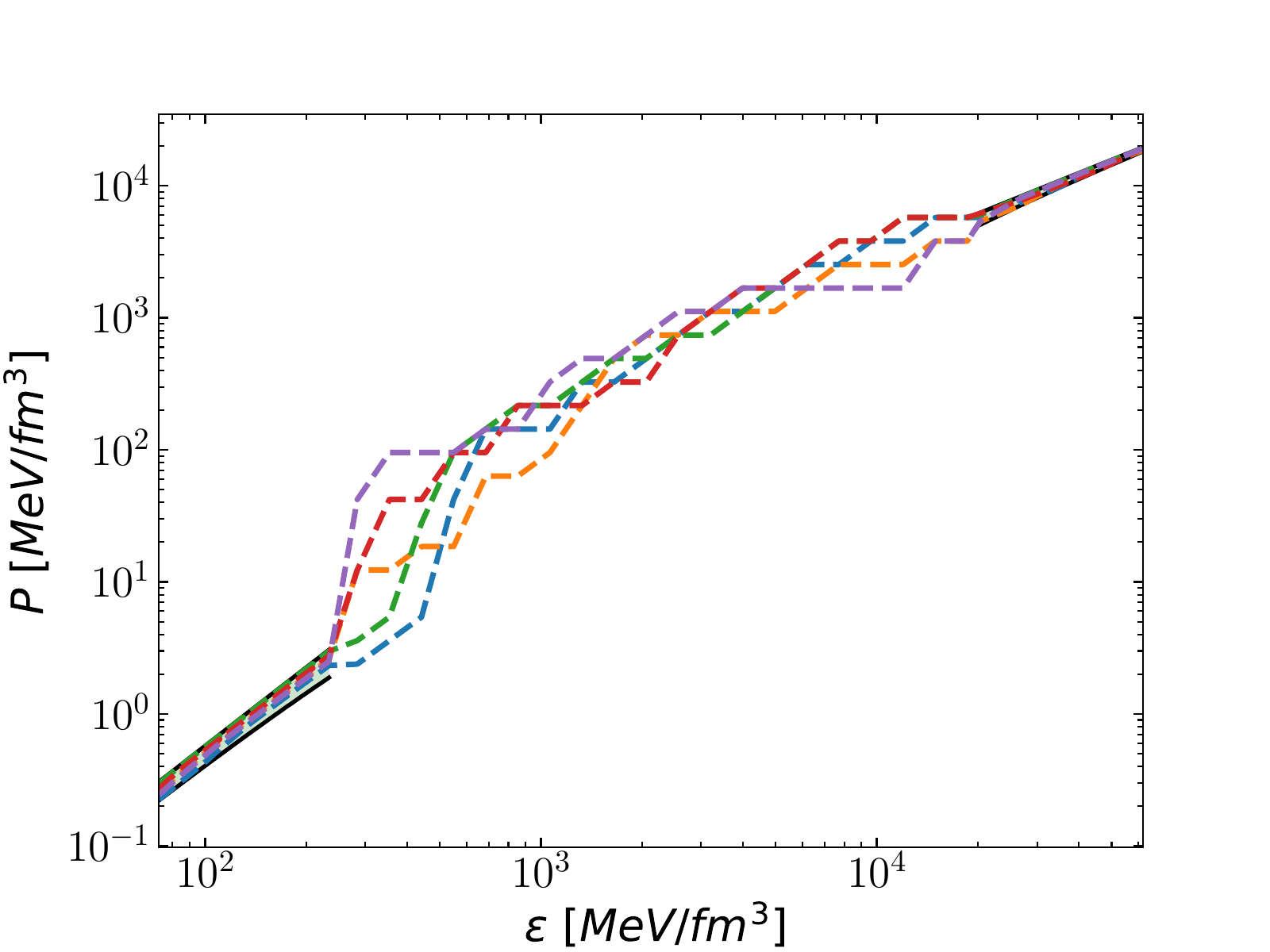}
   \end{minipage}
\caption{Several example EoS satisfying all constraints. \emph{(I)} The low--density band is constrained by NLO chiral potentials 
following Sammarruca \emph{et al.}~\cite{Sammarruca:2014zia}  with momentum cutoff $\Lambda=450$ and 600 MeV  (left panel) and  by Holt \emph{et al.}~\cite{Holt:2016pjb} (right panel) with cutoff at 450 and 500 MeV. \emph{(II)} At intermediate densities only monotonicity and causality are imposed. \emph{(III)} The high--density band is the pQCD constraint from Kurkela \emph{et al.}~\cite{Kurkela:2014vha} starting at baryon chemical potential $\mu_{\rm match}$=2.8 GeV. 
\label{fig:potquiral1}
}
\end{figure}

\begin{figure}
   \begin{minipage}{0.46\textwidth}
    \includegraphics*[width=3.44in]{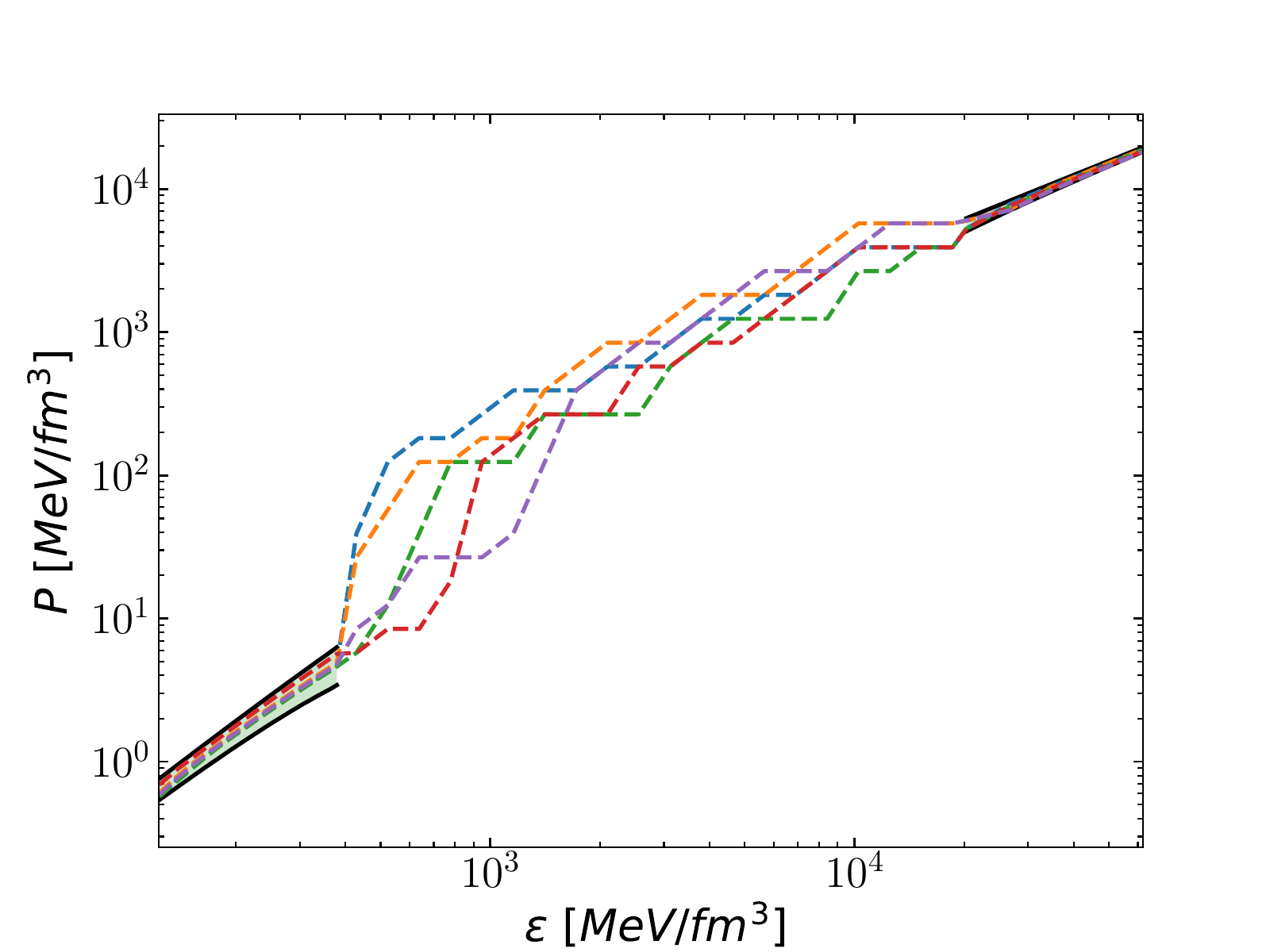} 
    \end{minipage}\hspace{0.5cm}
    \begin{minipage}{0.46\textwidth}
     \includegraphics*[width=3.44in]{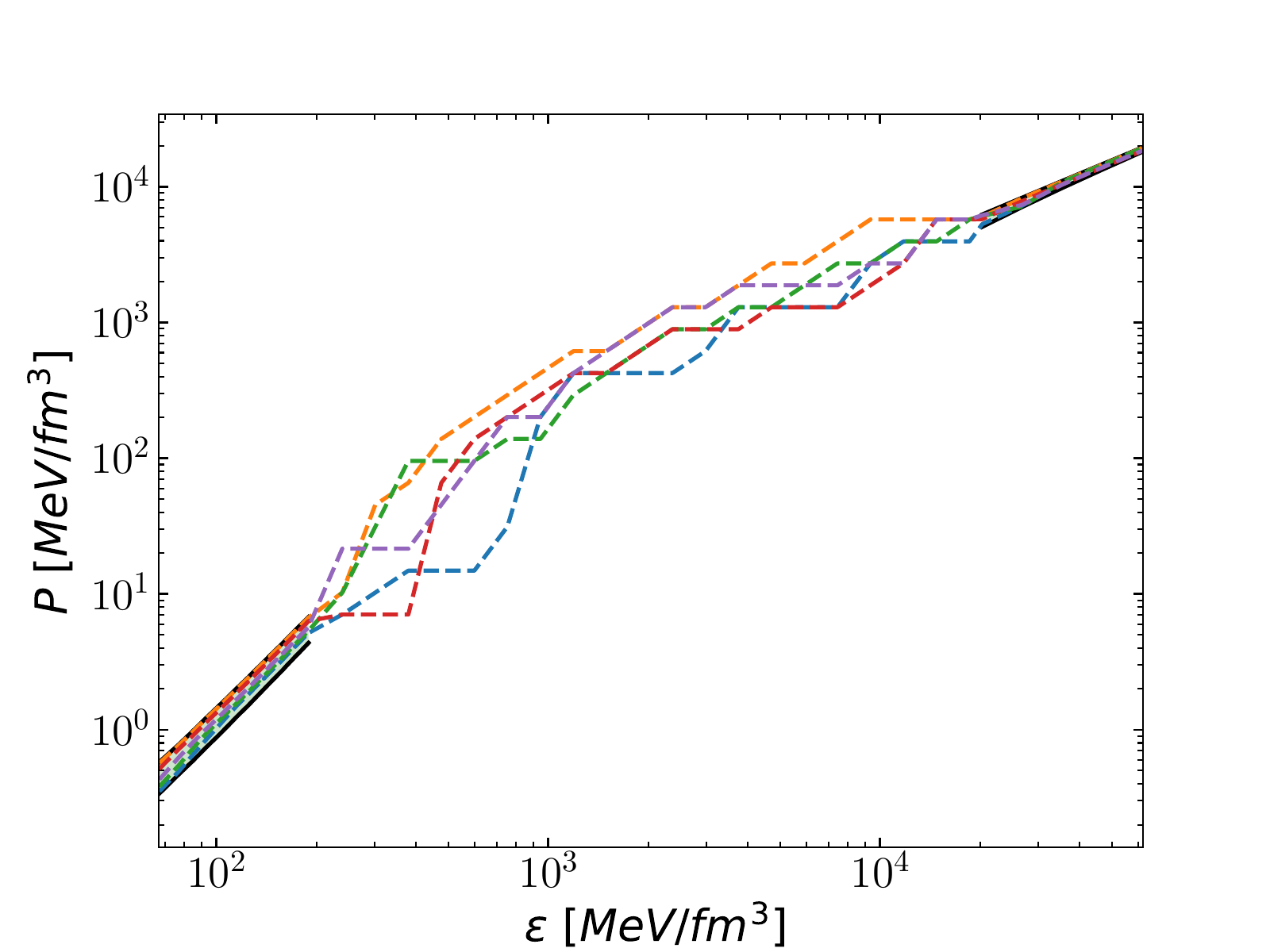}
   \end{minipage}
  \caption{Further example EoS as in figure~\ref{fig:potquiral1}
but with different the low--density bands of EoSes.
Left panel: N$^3$LO chiral potential for R=0.9 fm and R=1.0 fm from Hu \emph{et al.}~\cite{Hu:2016nkw}. Right panel: ChPT EoS from Drischler  \emph{et al.}~\cite{Drischler:2016djf}. 
 \label{fig:potquiral2}
}
\end{figure}

In figures~\ref{fig:potquiral1} and~\ref{fig:potquiral2} we show some examples of equations of state obtained following this procedure.
These example EoSes span a band in $(\varepsilon,P)$ space that is broader than in~\cite{Kurkela:2014vha} precisely because
we have imposed no observational constraint from astrophysics.

\section{Software package and sampled EoS} \label{sec:public}

We have set up a website to publicly distribute  this nEoS project at \\
\url{http://teorica.fis.ucm.es/nEoS/} ; 
the website provides a ``quick--start'' menu with a few typical EoS constrained by alternative chiral potentials (\emph{viz.} those from figures~\ref{fig:potquiral1} and~\ref{fig:potquiral2}, classified by the order of the chiral potentials from NLO up to partially N$^4$LO), with the interaction cutoff at two different momenta when available, and with the matching point to pQCD slightly varied at 2.6 or 2.8 GeV alternatively, to test the sensitivity to this parameter. This serves the purpose of code testing and can be useful for illustration purposes in various contexts. 
In this section we will use a sampler of these ``quick--start'' tables of the pair $(\varepsilon,P)$ to informally show what the user should expect of the nEoS sets.

We have also provide tables in {\tt .tar.gz} files containing a larger sample of EoS for use in production work. They are again classified according to the chiral potential employed at low density (all of them match the essentially unique high--density computation of pQCD), and by the momentum cutoff up to which the ChPT computation is used.

A delicate choice is how to set the density of points in the grid: an excessive number of points will require much time to interpolate the EoS for possible future massive computational applications in general relativity and modifications thereof. (Practitioners there are often used to very fast polytrope EoS parametrizations and it is not clear how well the interpolated EoS will perform with a variety of computer codes.)

On the other hand, a low-density grid limits the sample to a small number of distinguishable equations of state (as there is a limited number of $(\varepsilon_i,P_j)$ pairs that can be sampled). We have opted for a grid with modest size, about 50 points along the $\varepsilon$ axis (logarithmically distributed between $2.6\times 10^2$MeV/fm$^3$ and $4.3\times 10^5$MeV/fm$^3$). This is enough to allow for flat plateaux that represent first-order phase transitions, as seen in the samples shown in figures~\ref{fig:potquiral1} and~\ref{fig:potquiral2}, and to reach high enough energy density, above (3 GeV)$^4$, to allow credible matching to pQCD.

On the negative side, this choice of a relatively sparse grid constrains the number of different EoS that can be constructed to 
$O(10^5)$. To reduce to a minimum the probability that any two of the EoS that we generate and provide can be equal, we have limited the number to $10^3$ for each of the sets.  The probability of repeating an EoS in the sample is then very small (a short script can be executed to check for repetitions).

\section{Tests of the generated Equations of State}
\label{sec:tests}

We have solved the Tolman-Oppenheimer-Volkoff equations of hydrostatic equilibrium to provide a feeling for these bands of EoS, 
choosing one of the sets of example EoS (that with a low--density limit taken from~\cite{Drischler:2016djf}.

\begin{figure}
\centering
    \includegraphics*[width=3.5in]{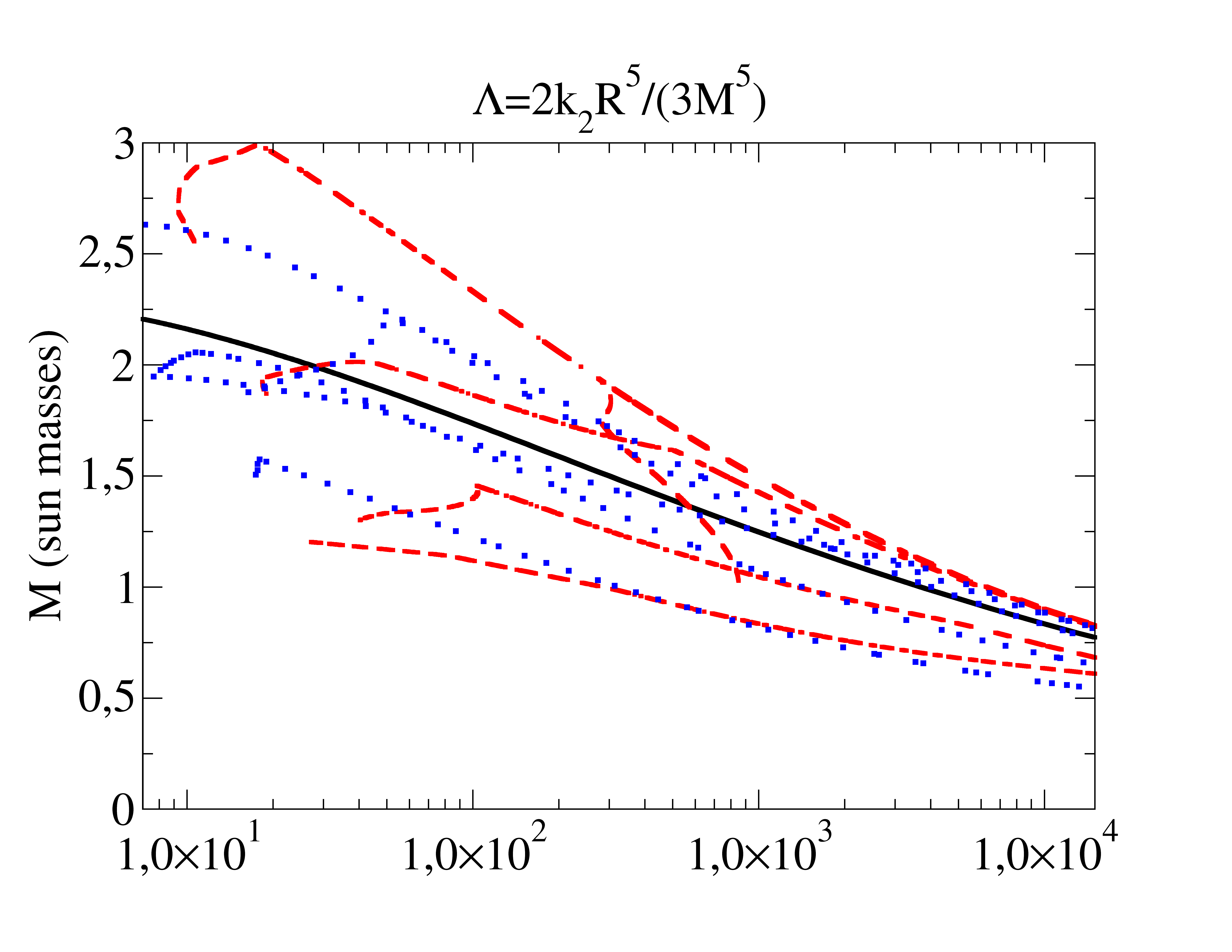} \\
     \includegraphics*[width=3.5in]{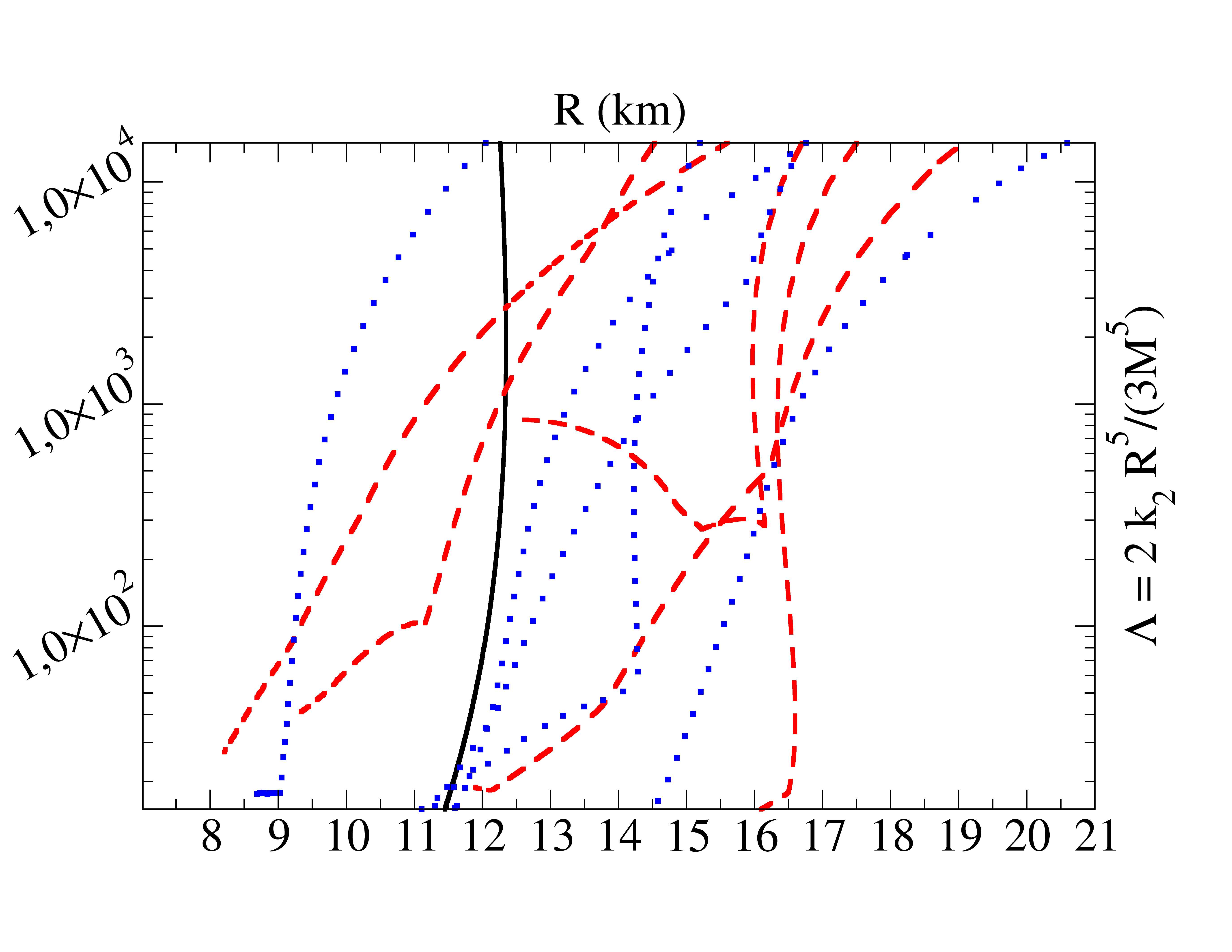}
\caption{
Dimensionless tidal deformability against the neutron star mass (top) and the neutron star radius (bottom) for a selection of EoS interpolated between the low density chiral perturbation EoS from~\cite{Drischler:2016djf} and the high-density pQCD EoS from~\cite{Kurkela:2009gj}. Matching to pQCD occurs at the pressure computed from that approach at $\mu_{\rm match}=2.6$ GeV or 2.8 GeV for the dashed lines (red online) and dotted lines (blue online) respectively. For comparison, we also plot a computation along the lines 
described in~\cite{Dobado:2011gd} with an EoS provided from~\cite{Lacour:2009ej} (solid line).
\label{fig:tidalDrischler}
}
\end{figure}

\begin{figure}
\centering
\includegraphics[width=4in]{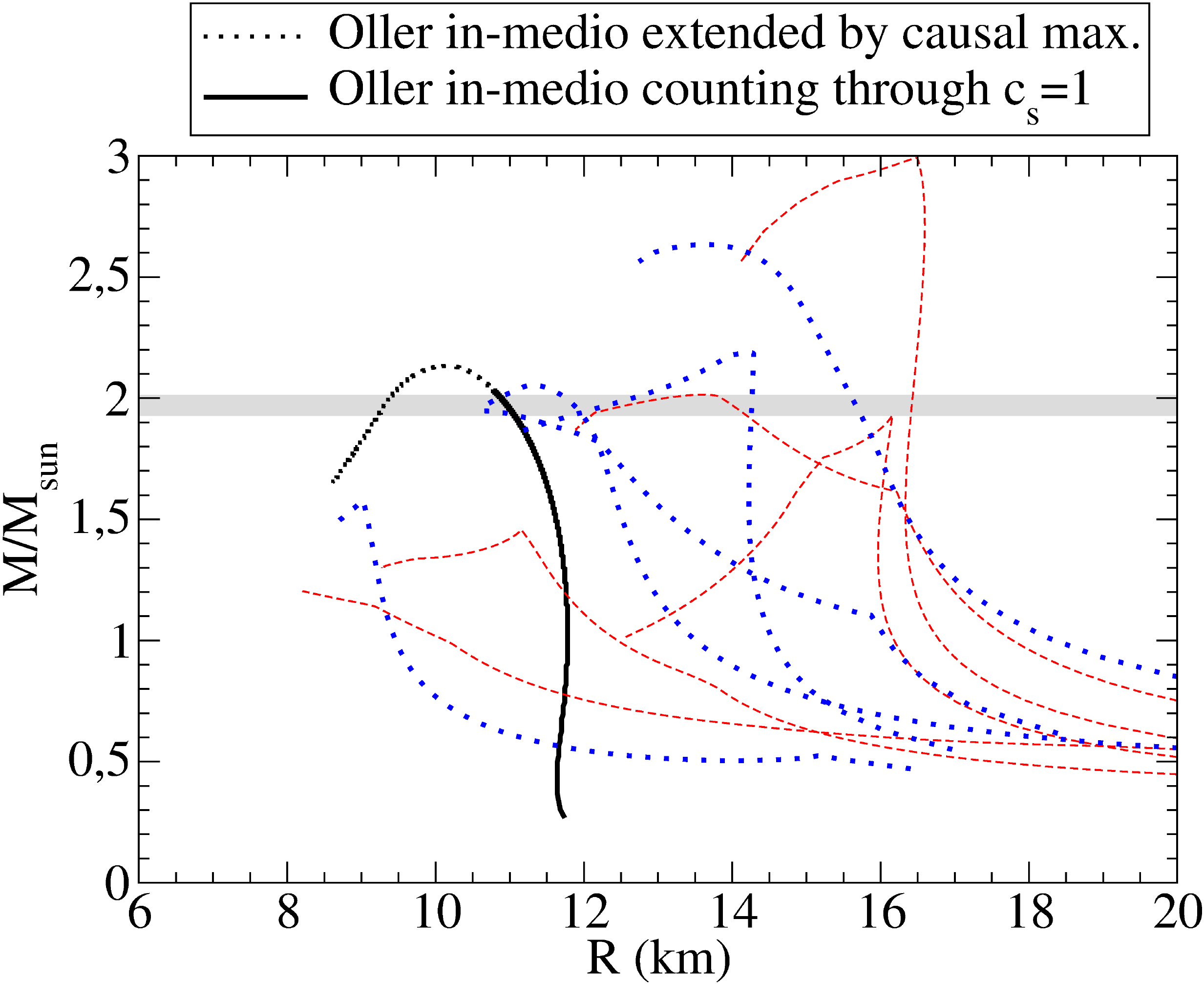}
\caption{\label{fig:MR} Standard mass--radius diagram for the same selection of EoS as in figure~\ref{fig:tidalDrischler}.
}
\end{figure}

The resulting tidal deformability~\cite{Hinderer:2009ca} and mass--radius diagrams are plotted in figures~\ref{fig:tidalDrischler} and~\ref{fig:MR}. While the absolute deformability $\lambda$ is the proportionality constant between the quadrupole tensor deformation of the star $Q$, and the tidal field causing it, $\mathcal{E}$, \emph{i.e.} $Q_{ij}=\lambda \mathcal{E}_{ij}$, it has become customary to normalize it differently. Following the aLIGO-VIRGO discovery of GW170817, the literature has been skewing towards the dimensionless quantity $\Lambda=\frac{\lambda}{M^5}$, and this is what we plot in figure~\ref{fig:tidalDrischler}. (the gravitational wave data constrains the mass-weighted tidal deformability $\tilde{\Lambda}$ that is a property of the binary system, not of each individual star; when $M_1=M_2$ in the binary system, $\tilde\Lambda\equiv\Lambda$).

The dimensionless tidal deformability varies across a wide range. 
It is quite uncorrelated from the neutron star mass and radius, so that it can be thought of as a third, independent observable: in GR this can be used to further constrain the EoS. It would be interesting to repeat the computation of the tidal deformability in modified gravity along the lines of~\cite{Yazadjiev:2018xxk} with the nEoS sets to see how broad the swath of lines is and whether a constraint on the allowed separation from Einstein's theory is possible with the full nEoS sets or, on the contrary, the maximum mass of a neutron star remains the more constraining observable.

For comparison we also added the computation of~\cite{Dobado:2011gd} with the in-medio counting of~\cite{Lacour:2009ej} (but for which no uncertainty band has been yet calculated). 

The two sets shown (with dashed and dotted lines respectively; they correspond to sets 7a and 7b in
our collection at \url{http://teorica.fis.ucm.es/nEoS}) have slightly different matching baryon chemical potentials
$\mu_{\rm match}=2.6$ and 2.8 GeV respectively, for the transition to pQCD matter. The sensitivity to this choice is small unless one lowers $\mu_{\rm match}$  to significantly lower values, say 2 GeV; then the pQCD EoS becomes quite steep (unlike the slow logarithmic running for higher densities) and this affects the interpolated intermediate energy density.

The plotted lines look somewhat more ragged than usually reported computations; this is because of the nonsmooth EoS in the set (as we are not employing soft polytrope--like interpolation at all), so the various curves occasionally change direction.

The basic feature of a maximum neutron star mass below three solar masses is there, which can help constrain variations of General Relativity. The differences among computed radii (and consequently among tidal deformabilities) are larger than thought but we hope upcoming theory efforts will reduce this. The radius-independent quantities such as the Love number $k_2$ are much less-sensitive. It is straightforward to extend these examples with the 
EoS listed  in~{\tt http://teorica.fis.ucm.es/nEoS/} or to proceed to systematical investigations with the larger sets with 1000 equations each, also stored there.

\section{Conclusions} \label{sec:conclusions}

\subsection{Results}

In this concise work we have reported an extensive set of $\beta$-equilibrated neutron--star matter  Equations of State
based on first principles, on Chiral Perturbation Theory and on perturbation theory in QCD.
In addition to avoiding systematic model biases as much as currently possible, we have not
imposed any astrophysical constraints, so that the family of EoS truly samples the current
state of the art \emph{hadron theory prediction}.

This means that uncertainty bands for observables calculated from our nEoS family of equations
will be broad, as illustrated with the examples in figures~\ref{fig:tidalDrischler} and~\ref{fig:MR}, and we look 
forward to theory improvements to reduce these. 

We provide in all 18000 equations of state that satisfy all theory constraints, parceled in subsets of 1000 EoS each. 
These are organised as follows. 
\begin{itemize}
\item Sets 1 through 3: low density band constrained by~\cite{Sammarruca:2014zia} at NLO, N$^2$LO and N$^3$LO respectively, spanning cutoffs of 450 and 600 MeV. Provided as two subsets each, one with $\mu_B=2.6$ GeV (subsets 1a, 2a and 3a) and another with 2.8 GeV (subsets 1b, 2b, 3b).
\item Sets 4 through 6: low density band constrained by~\cite{Hu:2016nkw} at NLO, N$^3$LO and N$^4$LO, with cutoff parameters $R=0.9$ and 1 fm. Also provided as two subsets each, likewise denoted by (a) and (b) lettering.
\item Set 7: low density band constrained by~\cite{Drischler:2016djf}, subsets 7a and 7b as above depending on the $\mu_B$ matching to pQCD.
\item Sets 8 through 10: low density band constrained by~\cite{Holt:2016pjb}, at NLO, N$^2$LO, N$^3$LO respectively and also divided in (a) and (b) subsets each.
\end{itemize}

A systematic comparison of all the sets with respect to several simple observables is envisioned for the near future, but it may be that a different collaboration will want to undertake the endeavor, perhaps more efficiently than us, since the automation of neutron star calculations is a subfield in itself~\cite{Most:2018hfd}, so we have opted for releasing the sets for public examination.

\begin{figure}
\centering
  \begin{minipage}{0.49\textwidth}
    \centering
    \includegraphics*[width=3.in]{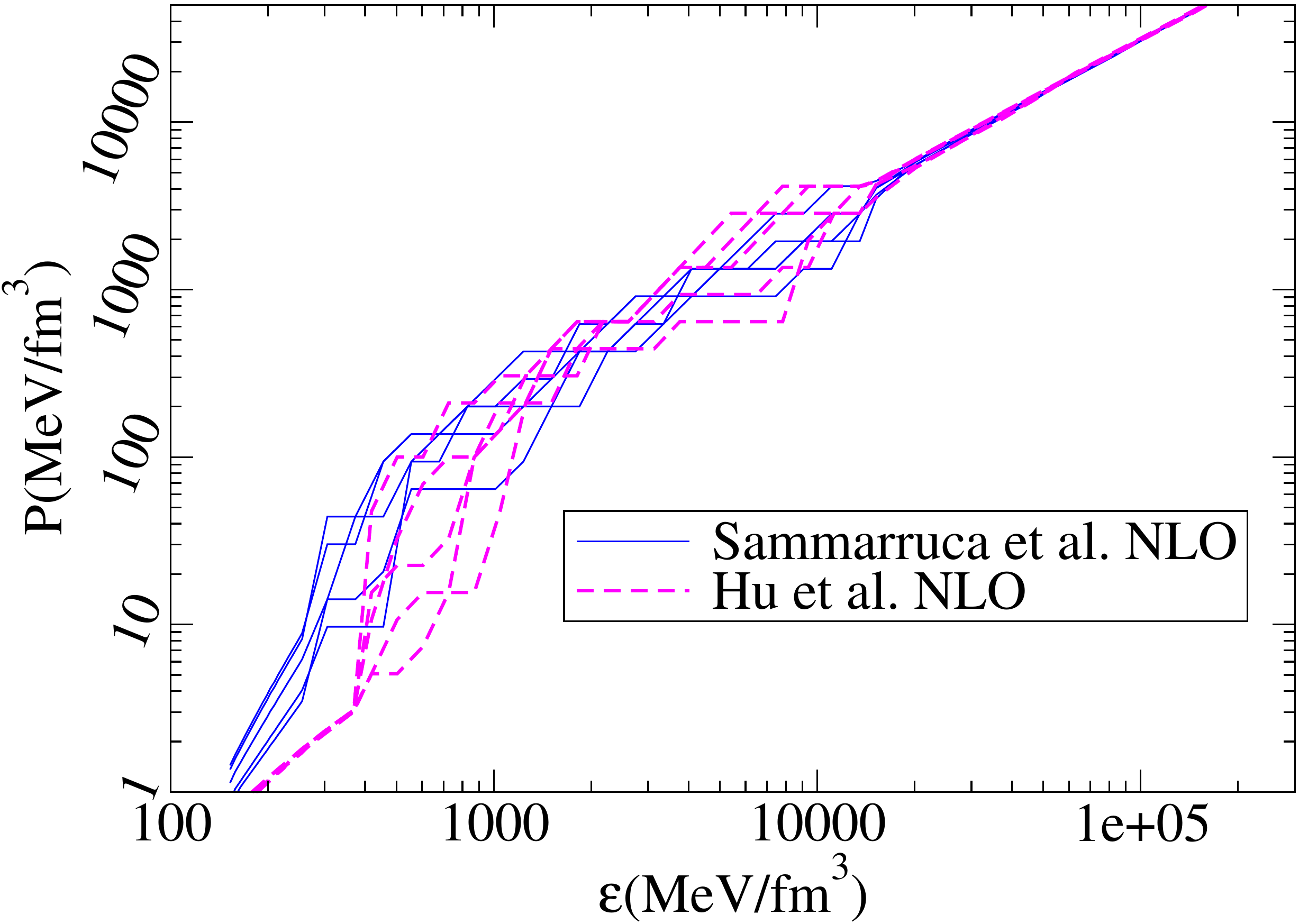} 
    \end{minipage}
    \begin{minipage}{0.49\textwidth}
    \centering
     \includegraphics*[width=3.in]{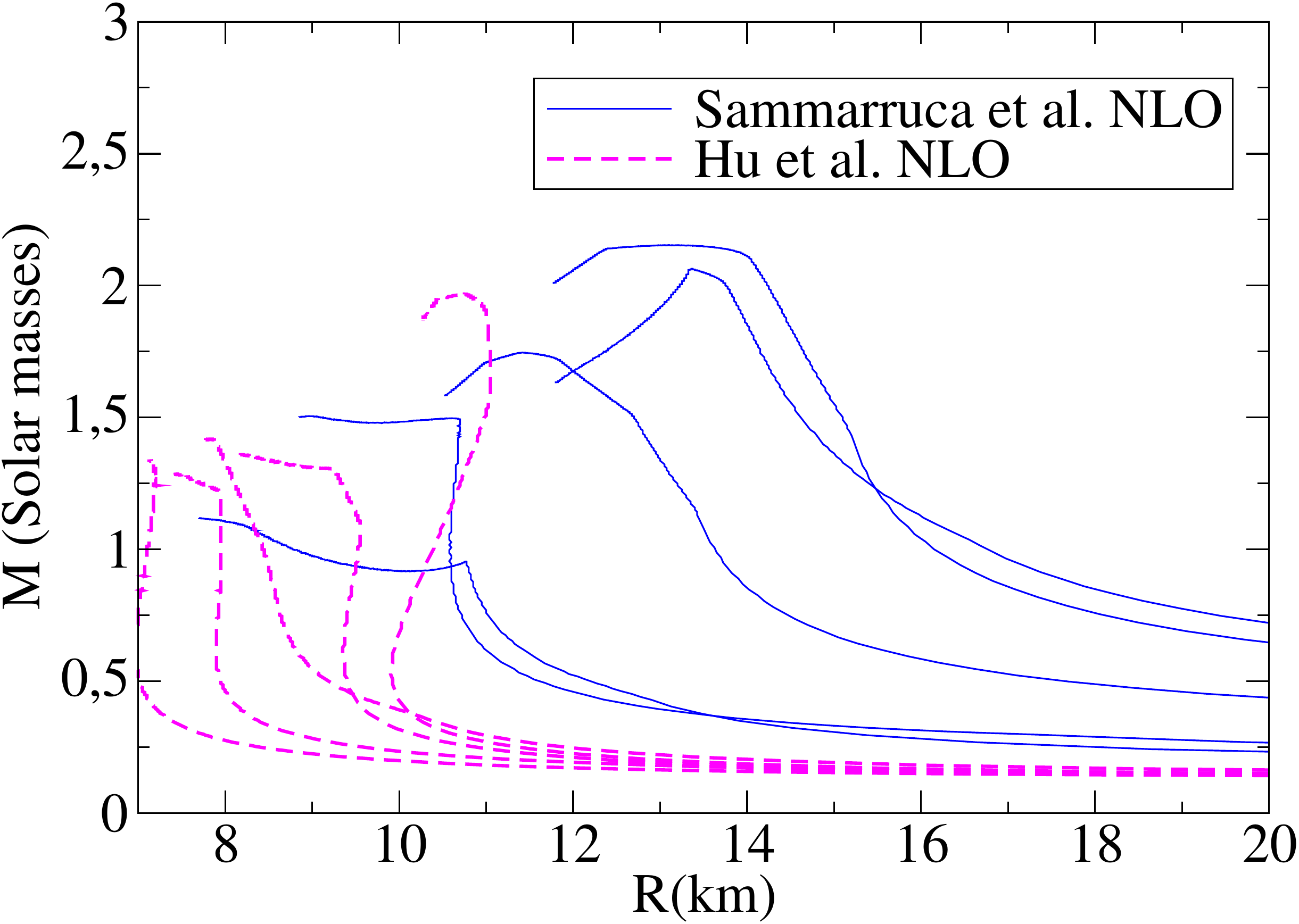}
   \end{minipage} 
\caption{Left plot: the five example EoS from subset (1a) constrained at low--density by~\cite{Sammarruca:2014zia} (Sammarruca \emph{et al.}) and the example EoS from subset (4a) constrained by~\cite{Hu:2016nkw} (Hu \emph{et al.})  Right plot: mass-radius diagram obtained from those 10 EoS. 
\label{fig:compareNLO}
}
\end{figure}

For an example of such work, let us plot in figure~\ref{fig:compareNLO} a comparison between two same--order subsets, namely (1a) and (4a) in the nEoS site, both using NLO input from chiral computations from different collaborations~\cite{Sammarruca:2014zia,Hu:2016nkw}. 

We can appreciate in the figure's left plot, the rendering of the nEoS sets based on (1a) (Sammarruca {\it et al.}) and (4a) (Hu {\it et al.}) how the first starts at higher pressure, so that the low--density band only reaches up to around 200 MeV/fm$^3$ in energy density, while the second, that is initially less steep, has a low--energy band extending up to 375 MeV/fm$^3$.  (The low--density band becomes doubtful when the typical neutron momentum approaches the cutoff).   At intermediate densities, both subsets start looking very similar, but subset (4a) never regains the stiffness lost at initial energy densities.

In consequence, subset (1a) reaches higher neutron--star masses and larger radii for a typical $1.4M_\odot$ neutron star: we find $R_{1a}\simeq (12.8\pm 2.1)$km, $R_{4a}\simeq(9\pm 1.8)$ km respectively.
A lot of predictive power can be gained \emph{from hadron physics alone} if the low--density investigations of the EoS in Chiral Perturbation Theory converge to a solidified result.

\subsection{Discussion and outlook}

There are several ways in which this project could be enhanced.

One aspect would be to incorporate results from lattice gauge theory. We have not done this yet because currently lattice gauge calculations have to be extrapolated from zero baryon density by using a Taylor expansion around $\mu_B=0$ and assuming analyticity in the baryon chemical potential~\cite{Borsanyi:2012cr}. This may be reliable for small $\mu_B$ at finite temperature, but still relies on a further model--dependent Hadron Resonance Gas extrapolation to reach zero temperature from the actual lattice calculations at $T\sim 100$ MeV.  The method can well fail at higher densities, where singularities might appear, particularly near a phase transition. Additionally, the region of parameter space covered, after all extrapolations, would coincide at best with that where we have effective (chiral) theory working. The only real gain by using lattice gauge theory would be to have the low-density EoS directly connected to $\alpha_s$, the strong coupling constant at the quark level, without resource to the coefficients of the effective Lagrangian; even this much looks like a far goal given the present status of work, that is making strides only for heavy quarks~\cite{Philipsen:2018rlz}.  In the practice of ChPT, the coefficients are fitted to $NN$, $\pi N$ and other low--energy data in laboratory experiments. For our purposes, since both approaches, QCD computation of the chiral coefficients, or accelerator data fitting,  are independent of General Relativity, it does not make a difference as long as the Equation of State becomes known. We will probably revisit this decision at a later date when/if progress in lattice gauge theory brings significant advantage.

A second caveat worth mentioning is that we have not incorporated the outer crustal EoS, described by (significantly messy) nuclear phases with drops of actual nuclei, perhaps ``pasta'' phases, as opposed to a simpler, fixed-$r$ homogeneous, $\beta$-equilibrated hadron continuum as described by chiral perturbation theory. 
It is well-known that this omission is of no concern for the mass of the star, as the crust is relatively much lighter, but it does bring--in an underestimate of the star radius by about 0.5 km~\cite{Fortin:2016hny}; but given that the uncertainty in experimental data is still about 2 km for the radius, and the spread of our EoS band gives easily a theory uncertainty of 4 km, which is eight times bigger, it is not urgent to incorporate it. 
We may do so at a future date if a way of providing a relatively model--independent estimate is pointed out to us. Meanwhile, we point the reader out to~\cite{Sharma:2015bna} for an appraisal of the nuclear--end EoS.

At the highest densities, it is expected~\cite{Alford:2007xm} that quark matter is in the Color-Flavor-Locked (CFL) phase, with Cooper pairing involving all three quark flavors. The pressure then contains a term proportional to the new scale, the pairing gap $\Delta$, of the form $P_2\propto \Delta^2 \mu_B^2$. This is subleading to the $\mu_B^4$ scaling of the ungapped pQCD computation in~\cite{Kurkela:2009gj} and is presently not included.

Finally, a compelling avenue for future work is to try to incorporate the effect of a
finite temperature into nEoS, which is necessary for neutron star mergers, perhaps using data from heavy ion collisions~\cite{Tsang:2018kqj}.

We are looking forward to addressing some of these and other interesting topics in future revisions of 
the EoS sets if they become of use for the community of researchers trying to test General Relativity with neutron stars.

\subsection*{Acknowledgments}
FJLE acknowledges the assistance of Iv\'an Sayago Galv\'an in the early stages of this investigation.
Work supported by grants MINECO:FPA2016-75654-C2-1-P (Spain); the EU COST Action PHAROS (CA16214); and by the U.S. Department of Energy, Office of Science, Office of Nuclear Physics under Award Number \#DE-FG02-05ER41375.
AW acknowledges support by the Schr\"odinger Fellowship J 3800-N27 of the Austrian Science Fund (FWF).

\newpage


\vspace{2cm}



\end{document}